# Role of Intrinsic Motivation in User Interface Design to Enhance Worker Performance in Amazon MTurk

Pushyami Kaveti*
kaveti.p@husky.neu.edu
Northeastern University
Boston, MA 02115

Md Navid Akbar*
makbar@ece.neu.edu
Northeastern University
Boston, MA 02115

## ABSTRACT

Biologists and scientists have been tackling the problem of marine life monitoring and fish stock estimation for many years now. Efforts are now directed to move towards non-intrusive methods, by utilizing specially designed underwater robots to collect images of the marine population. Training machine learning algorithms on the images collected, we can now estimate the population. This in turn helps to impose regulations to control overfishing. To train these models, however, we need annotated images. Annotation of large sets of images collected over a decade is quite challenging. Hence, we resort to Amazon Mechanical Turk (MTurk), a crowdsourcing platform, for the image annotation task. Although it is fast to get work done in MTurk, the work obtained is often of poor quality. This work aims to understand the human factors in designing Human Intelligence Tasks (HITs), from the perspective of the Self-Determination Theory. Applying elements from the theory, we design an HIT to increase the competence and motivation of the workers. Within our experimental framework, we find that the new interface significantly improves the accuracy of worker performance.

*Both authors contributed equally to this research

## 1 INTRODUCTION

Scientists have been collecting imagery of different types of fish and other marine organisms, for fish stock estimation and habitat monitoring. This has been going on for the last two decades. This is primarily achieved using remotely operated, autonomous underwater vehicles (AUVs) [1], which generally record images from a top view. Afterwards, the researchers manually go through each image trying to catalog information about the organisms they find in the images (mainly fish) and identify their species. This task involves manual labor, and thus requires a lot of human effort and time.

With the latest advancements in machine learning, a model can be trained to detect these underwater organisms. This way, the scientists can then focus on more important aspects like quality control, interpretation etc. Interestingly, to train a machine learning model, a large amount of labeled data is necessary. Researchers have this huge chunk of data, but their labels need to be accurately placed. This becomes a daunting task.

Labeling these huge datasets of underwater imagery via crowdsourcing (distributing the task among many people) is a promising solution [2]. Amazon Mechanical Turk (MTurk) is chosen as the platform to accomplish this, given its scalable nature. However, reliability and accuracy have emerged to be potential concerns in this sphere. There are many factors for these problems including diversity in workforce, variability in expertise and experience in the domain of the task, less attention span of the workers etc. Furthermore, there is very little control over the task after it is published as it is not a controlled laboratory-based work where feedback can be given to the worker. There is no dialogue or immediate interaction between the requestor and the worker due to technical difficulties. Thus, tasks on MTurk happen quickly, cheaply and with minimal interaction with workers. In this setting it is important to design HITs in way that they are simple, interesting and motivating. MTurk workers are motivated by various factors like money, enjoyment, interest, contribution to science etc. The goal of this work is to understand the motivating factors affecting the MTurk workers' performance in the HITs, and how we can to improve the HIT design to address the above-mentioned issues.

In this work, we apply Self-Determination Theory, which could explain the motivation behind people's actions to HIT design, in the context of crowdsourcing scientific image annotation. We aim to understand how this theory relates to the MTurk workers performance in HITs. This will help us to formulate favorable design rules for HITs. We propose hypotheses based on the Self-Determination Theory and design experiments to test the hypotheses. We then analyze the experimental results based on couple of performance metrics to assess the validity of our hypotheses. The following sections describe some of the previous work done and the details of the methodology and analysis.

## 2  PREVIOUS WORK

Crowdsourcing is increasingly used to develop large scale datasets for various scientific problems. Using MTurk to do image annotation on large datasets and methods used are described in [2,3]. As mentioned in these studies, the annotations need to be quality controlled to ensure accuracy. In Imagenet [3], multiple users annotate the same image and a voting scheme is used to choose the best label. In LSUN [4], an iterative process of annotation where labels for a small image subset obtained from MTurk workers are fed into a deep learning model to train and predict labels of a new set, some of which in turn are given to workers for annotation. This makes it interesting and important to understand the factors affecting the crowdsourced work quality and accuracy. There has been a lot of work done on understanding the characteristics of MTurk worker pool, why they become workers and how they perform [5]. In a survey of crowdsourcing systems [6], it is reported that demographics, financial incentives, intrinsic incentives like competence, learning etc., affect the performance aspect of workers. This is supported in [7], where researchers perform an analysis on monetary compensation as motivation and show that in India-based workers for whom MTurk can be a primary source of income, monetary incentive acts a motivation to do quality work.

Classic motivation theory and work motivation theory models are applied for MTurk in [8], and worker motivations are classified into extrinsic and intrinsic motivations. They show that intrinsic motivation factors like usage of skills, autonomy, enjoyment etc., dominate the extrinsic motivation factors like monetary incentives. Various intrinsic factors in crowdsourcing are studied, for example the use of task variety in improving MTurk worker motivation has been explored in [9]. The researchers in [9] found that variation in subtasks not only bolsters recruitment of more participants, but it also increases their retention and performance. Another interesting insight was discovered in [10], where researchers found that with a fixed payment scheme, workers usually did more work when no minimum work requirement was sought of them. These workers stated in their post-surveys that they continued to work, owing to their own interests. Thus, [10] concludes that task completion will observe a noticeable boost, if the tasks are engaging to the participants.

In our work we focus on two intrinsic motivation factors - competence and relatedness and evaluate their influence on MTurk workers. Related work is seen in [5], where qualification tests to screen out competent workers has been shown to be effective. Cognizance of contributing to a greater cause has also been suggested to improve worker motivation. This is particularly observed in the example, where workers were informed, they were finding tumor cells to help cure cancer [11]. Our work is inspired from these ideas. We aim to validate the effect of the intrinsic motivation factors mentioned above and support it with statistical evidence.

## 3  METHODOLOGY

In this section we describe our approach and implementation details of application of Self-Determination Theory to HIT design and evaluation.

### 3.1  Self-Determination Theory

This theory links human motivation and personality and involves the motivation behind the choices people make and things they do. It dictates that they are two types of motivation which shape who we are and what we do. They are

1. Intrinsic motivation: This corresponds to the internal drives that motivate people like interests, enjoyment, sense of morality, community etc.
2. Extrinsic motivation: This is drive that comes from external source like awards, respect from others, money etc.

In the context of MTurk, monetary compensation is the extrinsic motivation factor, which is shown to be dominated by intrinsic factors with respect to worker performance in HITs [8]. Thus, we choose to focus on two intrinsic motivation factors, competence and relatedness based on the nature of the task. The workers are asked to annotate marine species like rockfish, starfish and sponges in underwater imagery. These species appear very different in the dataset than how most of the people expect them to be, owing to their knowledge obtained from google search results or books. Understanding how the data actually looks like, and what and how to annotate, results in increased confidence and performance. Similarly, making the workers know the importance of the task and feel related also motivates them to perform better.

### 3.2  Hypotheses

We have developed two hypotheses built on the Self-Determination Theory. They are elaborated below.

Our first hypothesis targets the perceived competence and feeling of relatedness of MTurk workers. The competence factor may be introduced by providing a good tutorial to the workers and training them to perform the task better. The idea is that the MTurk workers will feel confident by knowing how to do the task, thus increasing their competence level. The relatedness factor is introduced by giving the MTurk workers a sense of belonging to the scientific community. We can provide workers with information about the importance of the task at hand and how it helps in solving a scientific problem. This will potentially give them a bigger picture and make them feel that they are contributing to the betterment of science and nature. From the obtained results of the surveys, we will investigate if the workers felt more competent and motivated while completing the enhanced HIT, compared to the basic HIT. We will test this hypothesis in Section 4.1.

We frame our second hypothesis around the actual performance of the workers in the HITs. Both the aforementioned factors have the potential to act as positive reinforcements and are expected to help people perform better on the enhanced HIT. Thus, we hypothesize that if the workers indeed felt more competent and motivated while completing the enhanced HIT, the accuracy of their annotations should also correspondingly increase. We will test this hypothesis in Section 4.2.

### 3.3 HIT Design

We developed two versions of the HIT. The first one is the control or basic HIT, which only has the instructions on how-to do the annotation as shown in Figure 1. The second version is the enhanced HIT, with design elements to incorporate motivation factors as shown in Figure 2 and Figure 3. The implementation details of design choices related to competence and relatedness of the enhanced HIT are given below:

1. We developed a guided practice test to teach the workers how to do the task. This is achieved in Amazon MTurk using a 'Qualification' test [5]. We designed this test using real underwater image samples, with annotations. Each question is followed with a set of multiple-choice options, among which only one is the correct answer. This test contains several edge cases, or extremes of different types of scenarios. A sample question in the practice test is shown in Figure 2. This way, the MTurk workers will obtain understanding of the manner in which the tasks are to be approached. By giving them a practice test, we attempted to increase their competence in doing the task accurately. Consequently, their resulting competence in differentiating between correct and incorrect ways of doing HITs should ramp up.
2. The relatedness hypothesis is implemented by including the purpose and significance of the annotation task in HIT. This is achieved by giving a brief description of the background information of the task and the overarching goal of preserving healthy populations of certain marine species, as shown in Figure 3. By briefing them, we are trying to make them understand the noble cause of their contribution to science and make them feel related to the community.

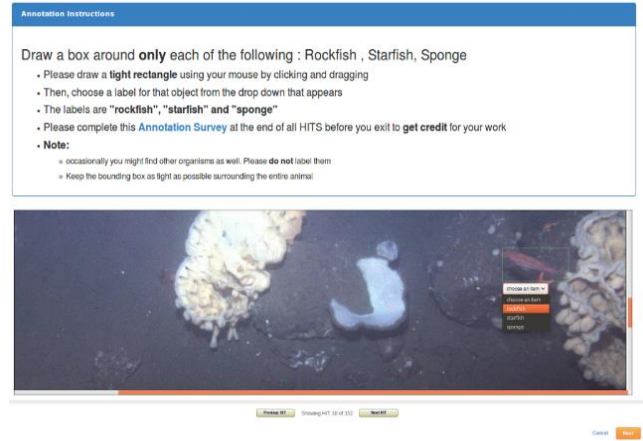

**Figure 1: Screenshot of the basic HIT design.**

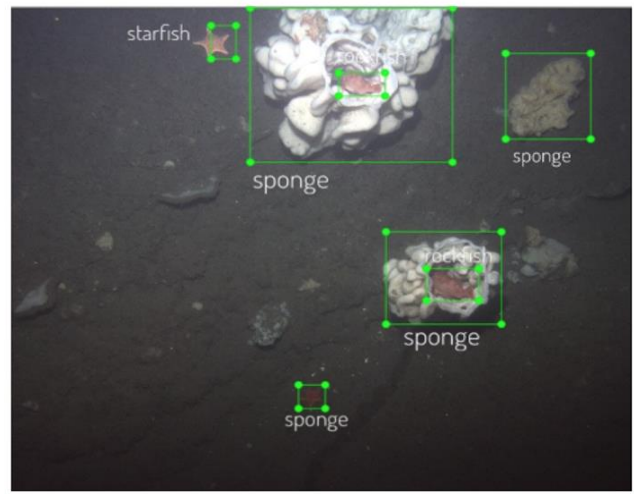

**Figure 2: A sample question from the practice test.**

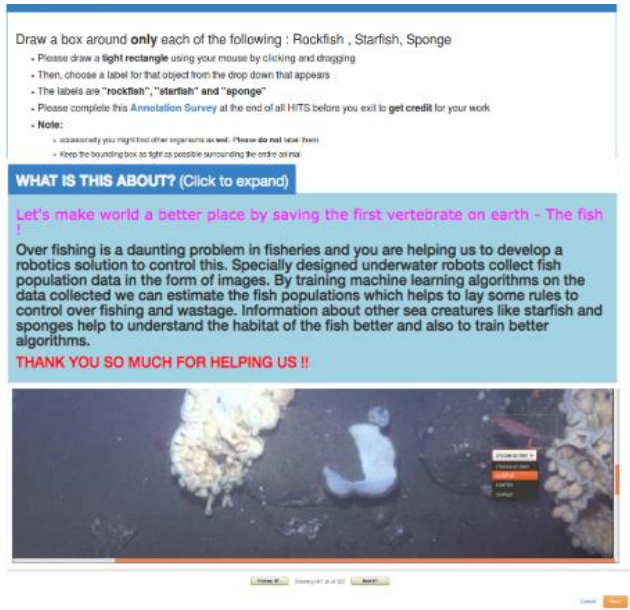

**Figure 3: Screenshot of the background motivational text.**

## 3.4 Experiments

After constructing these two HIT designs, we published them on MTurk in two sets. The first one, acting as a control, contained only the basic HITs, without the motivating texts or the practice test. The second batch was published with the motivating texts and the test. The guided test appeared as a qualifier in the worker site and sets them up for the HITs. The workers needed to take the test, and then qualify by scoring a certain minimum score to be able to proceed to complete the actual HITs.

Additionally, we included two comparable sets of post-completion survey links for the workers and instructed them to complete the survey before ending the last HIT to receive credit for their work. These surveys drew on the constructs of the Self-Determination Theory and implement elements from the Intrinsic Motivation Inventory. The surveys were intended to provide us insights into the perceived competence and motivation of the MTurk workers. The surveys were self-reported. They were kept short and concise, so that they do not confuse or stress the participants. Both the surveys were created using Google Forms, and stored in our secure accounts for analysis. The survey for the Basic HIT had the following statements:

a) I felt pretty competent while doing the HITs
b) This was an activity that I could not do very well. (R)
c) Rate your knowledge/skill in doing these annotations
d) I enjoyed doing these HITs very much.
e) I feel having a practice/tutorial will help me doing this task
f) I would like to know more about the background and why this task is important

The enhanced HIT survey had the following questions.
a) I felt pretty competent while doing the HITs.
b) This was an activity that I could not do very well. (R)
c) Rate your knowledge/skill in doing these annotations
d) I enjoyed doing these HITs very much.
e) I felt the Practice test did not help me at all. (R)
f) I felt the "What is this about?" description in HIT motivated me.

The survey answers were collected on a linear scale of 1 to 5, ranging from 'Strongly Disagree' to 'Strongly Agree'. These act as metrics to evaluate our first hypothesis. We also included a number of reversed questions, marked by (R). The purpose of this was to influence the workers to provide correct answers, as sometimes they just tend to glide through the questions, giving the same answer on the linear scale for each question.

The images used for annotation in this task were hosted on the Amazon Web Services (AWS) S3 bucket. The list of names of images are provided to MTurk via .csv file and the images are fetched by the MTurk platform to create all our HITs. We published 150 HITs for each of our experiments, where each HIT comprised of an image to be labeled.

After completion of the HITs, we assessed the survey results and accuracy of worker annotations as indicators of the performance metric. Ground truth labels were needed to calculate accuracy of labeling task and were generated manually by carefully labeling the 150 images using a tool called labelImg [12]. The accuracy of labeling is measured using two metrics:

1. Intersection over union (IOU): It is the ratio between the area of intersection to the area of union between the ground truth labels and the labels provided by MTurk workers.
2. Number of errors: It is the sum of false positives and unlabeled ground truths.

We utilized these metrics of accuracy to empirically assess the validity of our second hypothesis.

## 4 RESULTS

In this section, we will outline the results and test the corresponding hypotheses: of the self-reported survey responses and of the performance accuracy of the workers in the image annotation tasks. For both the tests, our hypotheses were tested as below:

| H0: There is no difference in the distribution of the two groups. |
|---|

| H1: There is a difference in the mean ranks of the two groups. |
|---|

### 4.1 Survey Responses

The self-reported survey questions were included in both sets of our HITs. The first four questions were identical for both the sets. Of these, we later selected two questions to ascertain the elements of competence and motivation of the MTurk workers. The final two questions in the survey sets were slightly different. In the first basic HIT, these questions were targeted to assess whether the MTurk workers felt the need for a practice test, and the background of the why the task is needed to be done. In the second enhanced HIT, these questions served to determine the efficacy of the practice test and the background information included in the enhanced HIT. The MTurk workers had five choices of ordinal responses: ranging between 'Strongly Disagree' (assigned a score of 1) to 'Strongly Agree' (assigned a score of 5). Since these are ordinal values, it has been suggested in past literature to use the mode for the response representation. The bar charts in Figure 4 below illustrates the survey results obtained for selected questions, from both the groups.

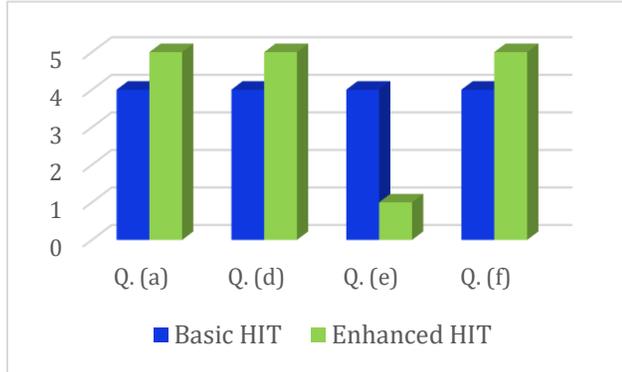

**Figure 4: Modes of selected survey question responses between the HIT groups.**

As can be observed from the modal responses to questions (a) and (d) in Figure 4, the MTurk workers who took the enhanced HIT reported higher levels of competence and enjoyment, compared to those who took the basic HIT. In by itself, the responses appear to reflect the effects of the practice test, and the motivational background text. However, this is not conclusive. Also, an additional challenge worthy of mention here is that we had lower number of survey responses, compared to the total number of annotation tasks. For the basic HIT, we had 118 responses ($n_1$), while for the enhanced, we had 64 ($n_2$). So, we went a step ahead, and applied the non-parametric Mann-Whitney U test on these variables. This test is specifically suitable for analysis of similarity and location of non-normal distributions of two distinct groups, with different group members, which is the case here. The results are summarized below in Table 1 and Table 2.

**Table 1: Mann-Whitney analysis for Q. (a) [Competence].**

|  |  |  |
|---|---|---|
| Basic | $R_1$ | 8982 |
|  | $n_1$ | 118 |
|  | $U_1$ | 5591 |
| Enhanced | $R_2$ | 7671 |
|  | $n_2$ | 64 |
|  | $U_2$ | 1961 |
| Combined Analysis | $U_{min}$ | 1961 |
|  | $U_{critical}$ | 3776 |
|  | $Z_u$ | -5.35 |
|  | $p$ | <0.00001 |

**Table 2: Mann-Whitney analysis for Q. (d) [Enjoyment].**

|  |  |  |
|---|---|---|
| Basic | $R_1$ | 8848 |
|  | $n_1$ | 118 |
|  | $U_1$ | 5725 |
| Enhanced | $R_2$ | 7805 |
|  | $n_2$ | 64 |
|  | $U_2$ | 1827 |
| Combined Analysis | $U_{min}$ | 1827 |
|  | $U_{critical}$ | 3776 |
|  | $Z_u$ | -5.74 |
|  | $p$ | <0.00001 |

In the above tables, $R_i$ is the sum of the response scores for the $i$-th HIT (*1* for basic, *2* for enhanced), and $U_i$ is the corresponding score from the Mann-Whitney U test. We then find out $U_{min}$ using

$$U_{min} = \min(U_1, U_2). \quad (1)$$

The fact that both $U_{min}$ were lower than their corresponding $U_{critical}$, it is indicative that there exists a difference between the locations of the two distributions, for each question. However, we do not know whether the difference is significant, or not. So, we proceed to find the corresponding $Z_u$ value. This is found using

$$Z_u = \frac{|U_{min} - (n_1 n_2/2)|}{\sqrt{\{n_1 n_2 (n_1 + n_2 + 1)\}/12}}. \quad (2)$$

This above relation uses a normal distribution approximation [13] and is valid when both the sample sizes are greater than 20. Using this value of $Z_u$, we then computed the value of $p$.

From the interpretation of the results in Tables 1 and 2, it can be concluded that the location of the distributions corresponding to the survey responses were different, at a two-tailed, 5% level of significance. So, the workers were indeed feeling more competent, and motivated while doing the enhanced HIT. Thus, we can reject the null hypothesis, and accept the alternate hypothesis. However, to be absolutely certain that these improved feelings of competence and enjoyment were really due to our intended changes, we investigated a little further.

For responses to the questions (e) and (f) In Figure 4, we notice that MTurk workers of the basic HIT felt the need for a practice test and why the task was important, marked by a high modal value of '4' in each. For the enhanced HIT, the MTurk workers reported negatively to our reversed question regarding the usefulness of the practice test, and positively to the significance of the motivational text, with a strong '1' and '5', respectively. This sets up our premise for two correlation analyses, which can confirm whether the aforementioned perceived feelings of competence and enjoyment could be attributed to our intentional changes.

We then performed a within group analysis for both sets of HITs. Here, we matched competence with the need or efficacy of the practice test, and enjoyment with the need or usefulness of the related background information. We calculated the correlation coefficient between these variables. Additionally, these variables are also represented as X and Y axes, to represent them in scatter plots. A least-squares regression line is then drawn for each plot, to visually indicate the relationship signified by the correlation values. These results are shown in Figure 5.

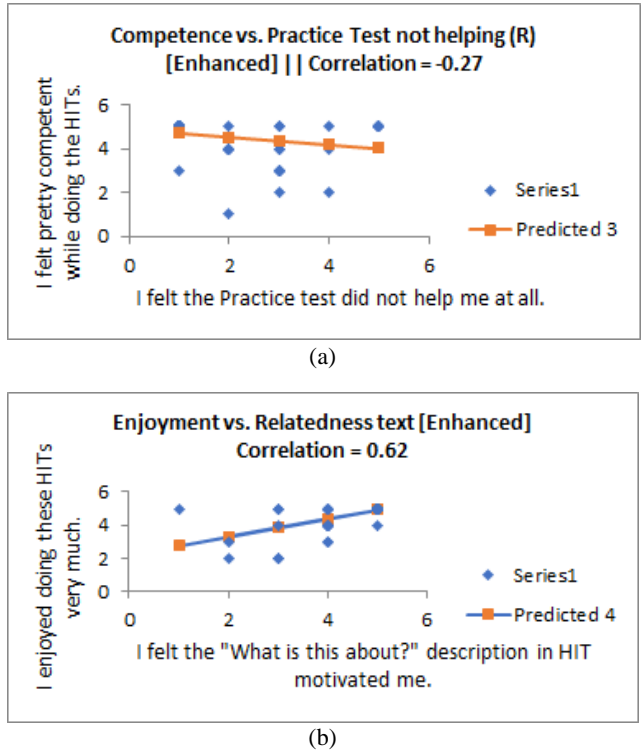

Figure 5: Correlation and regression analyses of the selected survey question sets, within the enhanced HIT.

Figure 5 (a) illustrates our first analysis. We observe that the correlation coefficient, $r_1$, is negative. This is our expected result, since we are comparing the competence, with the reversed question that the practice test did not help them at all. Thus, it provides evidence that the assumption behind our first hypothesis is reasonably justified. The value of $r_1$ is low, though. This indicates that the relation is not very strong. A possible explanation may rest in our aforementioned conjecture that some workers rushed their responses in this reversed question, without necessarily observing that it was worded differently.

Figure 5 (b) illustrates our second analysis. Here, we notice the correlation coefficient, $r_2$, is quite strongly positive. Again, this is our anticipated result, since we are comparing the enjoyment/motivation, with the effect of the background information to increase relatedness. The value of $r_2$ is also greater in magnitude, indicating a stronger relation between the variables. This analysis thus confirms that our first hypothesis is significant, within our experimental framework.

### 4.2   Image Annotation Tasks

We will now present the performance summary of the workers in the HITs. We first define the number of errors as

$Errors = Unlabeled\ ground\ truths + False\ positives.$   (3)

Figure 6 illustrates the distributions for the average IOU/HIT and error/HIT for each of the HITs, respectively.

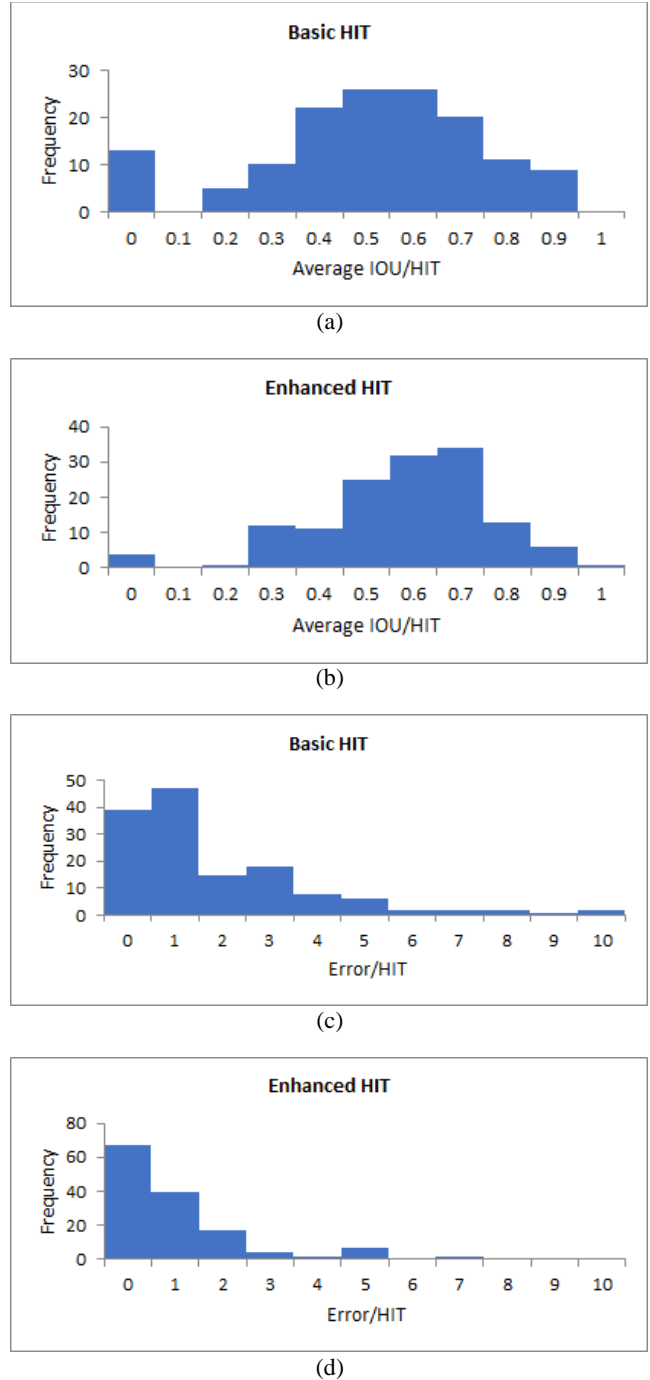

Figure 6: Histograms for the average IOU/HIT (a and b) and error/HIT (c and d) distributions, corresponding to each HIT.

As we can observe, these distributions are not Gaussian. Additionally, certain standard transformation on the data also did not yield satisfactory normal distributions. Consequently, we applied two-tailed Mann-Whitney U tests between the groups, for the average IOU/HIT and the error/HIT, with a 5% level of significance. Table 3 and Table 4 depict the results in detail.

**Table 3: Mann-Whitney analysis for average IOU/HIT.**

| | | |
|---|---|---|
| Basic | $R_1$ | 18327 |
| | $n_1$ | 142 |
| | $U_1$ | 8174 |
| Enhanced | $R_2$ | 21294 |
| | $n_2$ | 139 |
| | $U_2$ | 11564 |
| Combined Analysis | $U_{min}$ | 8174 |
| | $U_{critical}$ | 8534 |
| | $Z_u$ | 2.49 |
| | $p$ | 0.0124 (<0.05) |

**Table 4: Mann-Whitney analysis for error/HIT.**

| | | |
|---|---|---|
| Basic | $R_1$ | 22587 |
| | $n_1$ | 142 |
| | $U_1$ | 12434 |
| Enhanced | $R_2$ | 17035 |
| | $n_2$ | 139 |
| | $U_2$ | 7305 |
| Combined Analysis | $U_{min}$ | 7305 |
| | $U_{critical}$ | 8534 |
| | $Z_u$ | 3.77 |
| | $p$ | 0.0002 (<0.05) |

It may be noted that our sample sizes were 142 and 139, respectively, for the basic HIT and the enhanced HIT. Our original sample size was 150, each. We needed to filter out a few sets of data, less than ten percent of our sample size, due to errors in annotation. In these data sets, the MTurk workers drew multiple labels on the same marine animals. This caused IOU inaccuracies and produced anomalous results. To overcome this undesirable phenomenon, we discarded those samples.

Let us focus on the average IOU/HIT testing first. From Table 3, we notice that *p = 0.0124 (<0.05)*. This confirms the rejection of the null hypothesis. Thus, we accept the alternate hypothesis. This verifies that a difference in location of the two distributions exist. Moreover, the mean, median and mode of the IOU/HIT is higher in the enhanced HIT distribution, compared to those of the basic HIT. Thus, both of the above information taken in conjunction, we can affirm that the average IOU/HIT was higher in the enhanced HIT group.

We then shift our attention to the error/HIT testing. From Table 4, we find *p = 0.0002 (<0.05)*. This confirms the rejection of the null hypothesis. Thus, we accept the alternate hypothesis. This validates that there is indeed a difference in location of the two distributions. Moreover, the mean and mode of the error/HIT is lower in the enhanced HIT distribution, compared to those of the basic HIT. Thus, both above information taken together, we can conclude that the error/HIT was lower in the enhanced HIT group.

To sum up, we observed that both of our hypotheses have been justifiably validated from the survey results and the accuracy analysis of the HITs. Furthermore, when both the hypotheses taken together, they form part of the intrinsic motivation theory. This theory explains that by providing the workers with an opportunity to practice and making them feel related to the background of the task, their performance will boost. We have seen the average IOU/HIT to increase and the error/HIT to decrease, in the case of the enhanced HIT. The former is a metric of accuracy, while the latter is metric of correctness, and we have noticed improvements in both the metrics with the enhanced HIT.

## 5 CONCLUSION

The Amazon MTurk crowdsourcing platform is a great solution for annotating large image datasets, like underwater images discussed in this paper. However, this platform suffers from certain drawbacks, such as questionable quality of the work obtained. In this work we attempted to address that issue by the application of Self-Determination Theory to improve the competence and relatedness of the MTurk workers through Human Intelligence Task design. We came up with two hypotheses, which ties together competence with practice, and relatedness with motivation. With the analysis of the self-reported survey responses of the MTurk workers, as well as their performance in the HITs, we arrived at the conclusion that both of our hypotheses are indeed true, within our experimental settings.

## REFERENCES

[1] Clarke, M. Elizabeth, Nick Tolimieri, and Hanumant Singh. "Using the seabed AUV to assess populations of groundfish in untrawlable areas." *The future of fisheries science in North America*. Springer, Dordrecht, 2009. 357-372.
[2] Sorokin, Alexander, and David Forsyth. "Utility data annotation with amazon MTurk." *Computer Vision and Pattern Recognition Workshops, 2008. CVPRW'08. IEEE Computer Society Conference on*. IEEE, 2008.
[3] Deng, Jia, et al. "Imagenet: A large-scale hierarchical image database." *Computer Vision and Pattern Recognition, 2009. CVPR 2009. IEEE Conference on*. Ieee, 2009.
[4] Yu, Fisher, et al. "Lsun: Construction of a large-scale image dataset using deep learning with humans in the loop." arXiv preprint arXiv:1506.03365 (2015).
[5] Paolacci, Gabriele, and Jesse Chandler. "Inside the Turk: Understanding MTurk as a participant pool." *Current Directions in Psychological Science* 23.3 (2014): 184-188.
[6] Yuen, Man-Ching, Irwin King, and Kwong-Sak Leung. "A survey of crowdsourcing systems." *Privacy, Security, Risk and Trust (PASSAT) and 2011 IEEE Third International Conference on Social Computing (SocialCom), 2011 IEEE Third International Conference on*. IEEE, 2011.
[7] Kaufmann, Nicolas, Thimo Schulze, and Daniel Veit. "More than fun and money. Worker Motivation in Crowdsourcing-A Study on MTurk." AMCIS. Vol. 11. No. 2011. 2011.
[8] Litman, Leib, Jonathan Robinson, and Cheskie Rosenzweig. "The relationship between motivation, monetary compensation, and data quality among US-and India-based workers on MTurk." Behavior research methods 47.2 (2015): 519-528.
[9] Spatharioti, Sofia Eleni, and Seth Cooper. "On Variety, Complexity, and Engagement in Crowdsourced Disaster Response Tasks." Proceedings of the 14th International Conference on Information Systems for Crisis Response and Management. Albi, France. 2017.
[10] Spatharioti, Sofia Eleni, et al. "A Required Work Payment Scheme for Crowdsourced Disaster Response: Worker Performance and Motivations." English. In: Proceedings of the 14th International Conference on Information Systems for Crisis Response and Management. Ed. by FB Tina Comes and 14th International Conference on Information Systems for Crisis Response and Management. Albi, France. 2017.
[11] handler. D. and Horton, J.J. 2011. Labor Allocation in Paid Crowdsourcing: Experimental Evidence on Positioning, Nudges and Prices. In proceeding of: Human Computation, Papers from the 2011 AAAI Workshop, San Francisco, California, USA, August 8, 2011
[12] LabelImg toolbox . https://github.com/tzutalin/labelImg
[13] https://math.usask.ca/~laverty/S245/Tables/wmw.pdf